\documentclass[prl,english,showpacs,preprintnumbers,amsmath,amssymb,nofootinbib,twocolumn,superscriptaddress]{revtex4-1}
\usepackage{epsfig}
\usepackage{bm}
\pdfoutput=1

\usepackage[latin1]{inputenc}
\usepackage{graphicx}
\usepackage{color}
\usepackage{bbm}
\usepackage{amssymb}
\usepackage{amsmath}
\usepackage{tabularx}

\usepackage{dsfont}

\def\0#1#2{\frac{#1}{#2}}

\def\s0#1#2{\mbox{\small{$ \frac{#1}{#2} $}}}

\newcommand{\beq}{\begin{equation}}
\newcommand{\eeq}{\end{equation}}
\newcommand{\bea}{\begin{eqnarray}}
\newcommand{\eea}{\end{eqnarray}}

\usepackage{babel}
\makeatother
\begin{document}

\title{Pressure, compressibility, and contact of the two-dimensional attractive Fermi gas}

\author{E. R. Anderson}
\affiliation{Department of Physics and Astronomy, University of North Carolina, Chapel Hill, North Carolina 27599-3255, USA}

\author{J. E. Drut} 
\affiliation{Department of Physics and Astronomy, University of North Carolina, Chapel Hill, North Carolina 27599-3255, USA}

\date {\today}

\begin{abstract}
Using \textit{ab initio} lattice methods, we calculate the finite temperature thermodynamics of homogeneous 
two-dimensional spin-$1/2$ fermions with attractive short-range interactions. We present results for the
density, pressure, compressibility, and quantum anomaly (i.e. Tan's contact) for a wide range of temperatures 
and coupling strengths, focusing on the unpolarized case.
Within our statistical and systematic uncertainties, our prediction for the density equation of state
differs quantitatively from the prediction by Luttinger-Ward theory in the strongly coupled region of parameter space, but
otherwise agrees well with it.
We also compare our calculations with the second- and third-order virial expansion, with which they are in excellent agreement in 
the low-fugacity regime.
\end{abstract}


\maketitle

{\it Introduction.--~} 
Quantum mechanics in two spatial dimensions (2D) is a fascinating playground to understand fundamental 
physics in a wide variety of situations. It represents a necessary (though often odd!) crossover between the integrable 
systems that live in 1D and their much more challenging 3D counterparts. Interest in 2D ranges from
basic theory and experiment to marketable technological applications. 
Among the most salient features and systems in 2D we have: the peculiar behavior of Kosterlitz-Thouless phase transitions~\cite{BKT}; 
the possibility of understanding quark confinement analytically in certain models~\cite{confinement}; anomalous scale invariance in
non-relativistic systems~\cite{Hoffmann}; the challenge of high-temperature superconductors~\cite{HighTc}; and, of course, graphene~\cite{graphene}.

In recent years, experiments with ultracold atoms~\cite{RevExp,RevTheory} have made clear progress towards a clean and systematic
study of fermionic atom clouds in constrained or modulated optical traps, which closely resemble 
a 2D situation~\cite{Experiments2D2010,Experiments2D2011,Experiments2D2012,ContactExperiment2D2012,RanderiaPairingFlatLand,Experiments2D2014} 
(see also Ref.~\cite{Vale2Dcriteria} for a recent discussion approaching that limit). 
These systems have been
of singular interest due to their malleability: by now it is well known that the inter-atomic interaction can be tuned 
essentially at will, low-temperature degenerate regimes can be reached, spin- and mass-asymmetric systems can be studied,
and so on. As advances are thus made towards understanding the thermodynamics, structure and phases on
the experimental side, theorists are once again faced with the challenge of strongly coupled regimes, which defy 
analytic approaches.

Previous theoretical studies have characterized the crossover between Bose-Einstein condensation (BEC) and BCS pairing in 
2D via mean-field analyses~\cite{Miyake,BCSBEC2D,ZhangLinDuan}. The first determination of the ground-state equation of
state, however, appeared relatively recently in Ref.~\cite{Bertaina}, which used the diffusion Monte Carlo method.
Even more recently, 
Ref.~\cite{ShiChiesaZhang} performed an auxiliary-field quantum Monte Carlo study of the ground state where the
pressure, contact, and condensate fraction were determined.
At finite temperature, the equation of state was first computed in the virial expansion in Ref.~\cite{LiuHuDrummond},
and in the Luttinger-Ward approach in Ref.~\cite{Enss2D}.
Pair correlations were investigated using the virial expansion at about the same time in Ref.~\cite{ParishEtAl} and
in Ref.~\cite{BarthHofmann}. Collective modes were studied in Refs.~\cite{ChaffinSchaefer,EnssUrban,BaurVogt},
and the shear viscosity and spin diffusion in Ref.~\cite{EnssShear}.

In this work we offer a few essential pieces of the thermodynamic puzzle of attractively interacting fermions in 2D.
We implement a lattice Monte Carlo method to calculate the thermal and short-range correlation properties
of the system along the 2D analogue of the BCS-BEC crossover. 
Specifically, we determine the density equation of state, from which we extract the pressure and compressibility; all of these are experimentally measurable.
To our knowledge, the present is the first fully non-perturbative, {\it ab initio} calculation of the equation of state of this system
at finite temperature.
In addition, we present here the first calculation of Tan's contact parameter~\cite{TanContact,ContactReview,WernerCastin} 
at finite temperature in 2D that is free of uncontrolled approximations.

{\it Hamiltonian and many-body method.--~} 
The dynamics of dilute spin-$1/2$ non-relativistic fermions with short-range interactions is governed by
the following Hamiltonian:
\beq
\label{Eq:H}
\hat H \!=\!\! \int\! d^2x \! \left [ 
\sum_{s=\uparrow,\downarrow}\! \hat \psi_s^\dagger({\bf x})\!\left(\!-\frac{\hbar^2\nabla^2}{2m} \!\right)\! \hat \psi_s^{}({\bf x}) 
- g \hat n^{}_{\uparrow}({\bf x}) \hat n^{}_{\downarrow}({\bf x})
\right]
\eeq
where $\hat \psi_s^{\dagger}, \hat \psi_s^{}$ are the creation and annihilation operators in coordinate space for spin $s$ particles,
and $\hat n^{}_{s} = \hat \psi_s^{\dagger}\hat \psi_s^{}$ are the corresponding densities.

In a recent work~\cite{EoS1D}, we performed a similar study to the one presented here, but restricted to
1D. Here, we use the same methods (which are also closely related to those of Refs.~\cite{BDM1, BDM2, EoSUFG2} 
used in 3D) but applied in 2D. We placed the spin-$1/2$ Fermi system in a Euclidean 
spacetime lattice of extent $N^{2}_x \times N^{}_\tau$ with
periodic boundary conditions in the spatial directions and anti-periodic in the time direction. A Trotter-Suzuki factorization, 
followed by a Hubbard-Stratonovich transformation~\cite{HS}, allowed us to write the grand-canonical 
partition function as a path integral over an auxiliary field. That path integral was evaluated using 
Metropolis-based Monte Carlo methods (see, e.g., Refs.~\cite{Drut:2012md,QMCReviews}). 
Throughout this work, we use units such that $\hbar \!=\! m \!=\! k_\text{B} \!=\! 1$, where $m$ is the mass of the fermions.
The physical spatial extent of the lattice is $L\times L$, where $L= N^{}_x \ell$ and we take $\ell = 1$ to set the length 
and momentum scales.
The extent of the temporal lattice is set by the inverse temperature $\beta = 1/T = \tau N^{}_\tau$. The time step $\tau= 0.05$ 
(in lattice units) was chosen to balance temporal discretization effects with computational efficiency; in any case, those discretization 
effects are smaller than our statistical effects.

As in our previous study, the physical input parameters are the inverse temperature $\beta$, the chemical potential 
$\mu = \mu^{}_\uparrow = \mu^{}_\downarrow$, and the (bare, attractive) coupling strength $g>0$, where the latter determines 
the binding energy $\varepsilon^{}_\text{B}$ of the two-body problem (see below). From these, we form two dimensionless 
quantities: the fugacity $z = \exp(\beta \mu)$ and the dimensionless coupling $\beta \varepsilon^{}_\text{B}$. 
Using $z$ and $\beta \varepsilon^{}_\text{B}$ as parameters
facilitates comparisons with experiments, as well as with other theoretical approaches such as the Luttinger-Ward calculations
mentioned above.

The connection between the bare coupling $g$ and the 2D scattering length
$a_\text{2D}^{}$ via the two-body problem requires us to set the scales $\ell$ and $L$. To avoid ambiguities, 
we computed the binding energy $\varepsilon^{}_\text{B}$ of the two-body problem on each of the lattices studied, which allows us to 
characterize the coupling in terms of the dimensionless parameter $\beta \varepsilon^{}_\text{B}$ above.
This 2D case, in contrast to its 1D and 3D analogues, is peculiar due to its anomalous scale invariance.
Indeed, as may be inferred from Eq.~(\ref{Eq:H}), the dimensions of $\hat \psi_s$ are of inverse length (in the natural units
mentioned above), such that $g$ is dimensionless and therefore the system is classically (i.e. before accounting for quantum 
fluctuations) scale-invariant. The appearance of a two-body bound state with binding energy $\varepsilon^{}_\text{B}$,
upon turning on the interaction, is an example of a quantum anomaly: the classical scale invariance of $\hat H$ is broken by quantum effects.
Notably, massless quantum chromodynamics in 3D, also a scale-free theory at the classical level, displays a similar feature.

Lattice calculations of the kind we use are exact, up to statistical and systematic uncertainties. To address the former, 
we took 1000 de-correlated samples for each data point in the plots shown below, which yields a statistical uncertainty of order $3\%$.
The smoothness of our results show that those effects are indeed small.

To address the systematic effects, one must 
approach the continuum limit. Two-dimensional problems are computationally inexpensive relative to full 3D problems, but they are 
still challenging. In this first study, we restricted ourselves to $N_x^{} = 11,15,19$. The continuum 
limit is achieved by lowering the density while still remaining in the many-particle, thermodynamic regime.  In turn, this is 
accomplished by increasing the lattice parameter $\beta$, ensuring that the thermal wavelength $\lambda^{}_T = \sqrt{2 \pi \beta}$ 
satisfies $\ell \ll \lambda^{}_T \ll L $; at fixed $z$, this reduces the density.
In this work, we used $\lambda^{}_T \simeq 8.0$, which is well within the regime specified above. We then studied whether our 
results collapse to a universal curve 
when $\beta$ and $g$ are varied while $\beta \varepsilon^{}_\text{B}$ is held fixed.
Lattice sizes larger than $N^{}_x = 19$ are computationally more expensive but not impractical; however, we chose to fix that size 
and cover a wider region of parameter space instead.
Because our study proceeded at constant $\beta \varepsilon^{}_\text{B}$, increasing $\beta$ implies 
reducing $g$, which results in smaller uncertainties associated with the temporal lattice spacing $\tau$ in the 
Trotter-Suzuki decomposition; these are expected to be of order $1\!-\!2\%$ (see e.g. Ref.~\cite{BDM2}).  

%
\begin{figure}[t]
\includegraphics[width=1.0\columnwidth]{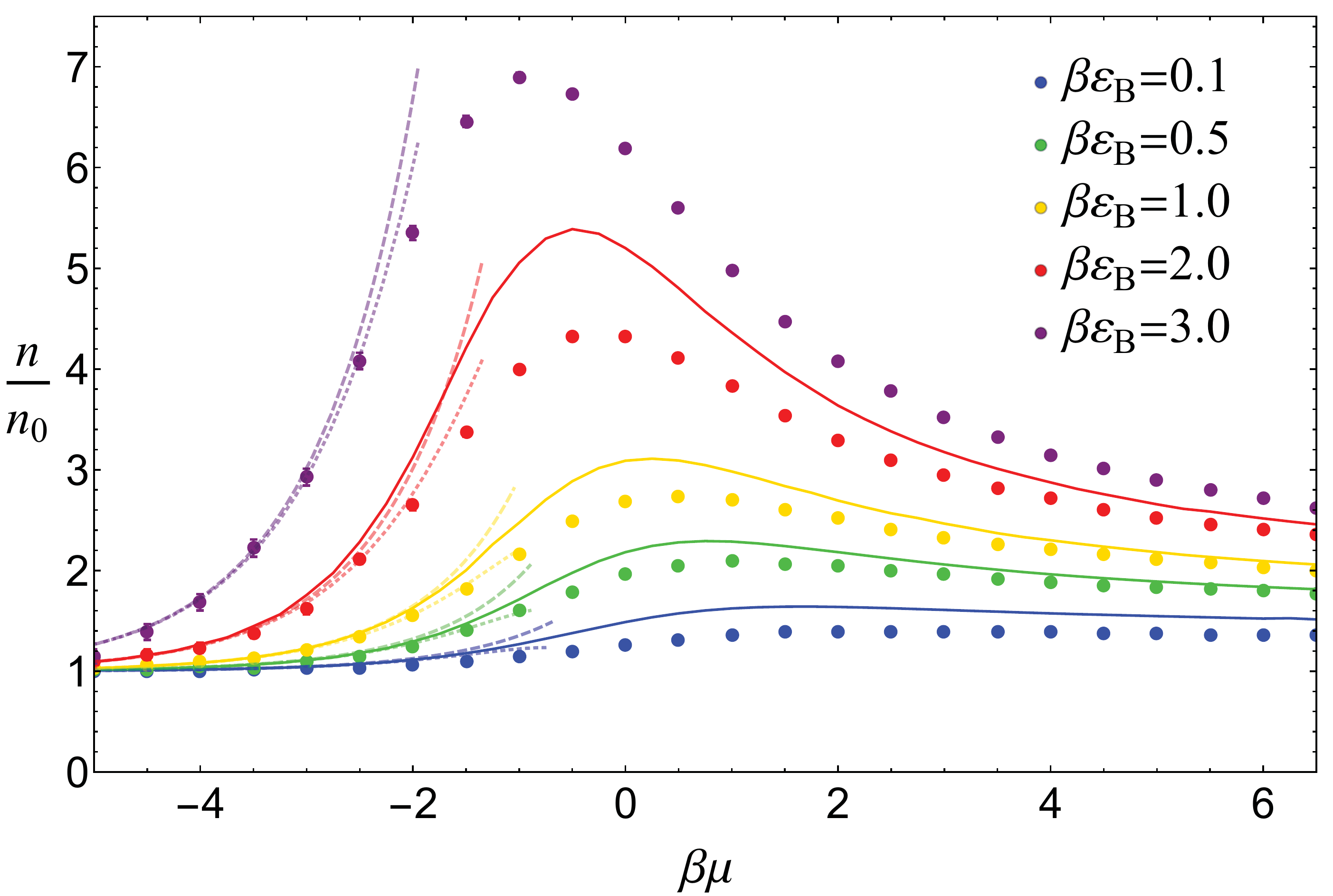}
\caption{\label{Fig:n_n0}(color online) Density equation of state, in units of the non-interacting density
$n^{}_0$ of spin-$1/2$ fermions in 2D, for coupling strengths $\beta\varepsilon^{}_\text{B}$ = 0.1, 0.5, 1, 2, 3 (from bottom to top), 
as a function of $\beta \mu$. The error bars reflect the statistical uncertainty.
The solid colored lines show the Luttinger-Ward result of Ref.~\cite{Enss2D}. 
The long- and short-dashed lines show the second- and third-order virial expansion results, respectively.}
\end{figure}
%

{\it Analysis and Results.--~} 
To characterize the thermodynamics of a strongly interacting system, as is our objective here, a simple yet 
extremely effective route is to first calculate the total particle number density $N/L^2$ (where 
$N = N^{}_\uparrow + N^{}_\downarrow$ and $N^{}_s$ is the particle number for spin $s=\uparrow,\downarrow$) as a function of the thermodynamic
variables (temperature, chemical potential, and interaction strength, as mentioned above). The density has the 
added benefit over other quantities that its statistical-noise effects in a Monte Carlo calculation are relatively small.
From $n$, one may determine the pressure $P$ by performing a numerical integration along $\beta\mu$, and the 
compressibility $\kappa$ by differentiation.

In Fig.~\ref{Fig:n_n0} we show the density $n$ as a function of the dimensionless parameters $z$ and $\beta \varepsilon^{}_\text{B}$,
defined above. The non-interacting result used as a scale in that figure is 
%
$
\label{Eq:n0}
n^{}_0 \lambda^{2}_T = 4 I^{}_1(z),
$
%
where $I^{}_1(z) = z\,{d I^{}_0(z)}/{d z}$, and 
\beq
I^{}_0(z) = \int_{0}^{\infty}dx\; x \ln(1 + z e^{-x^2}).
\eeq

\begin{figure}[b]
\includegraphics[width=1.0\columnwidth]{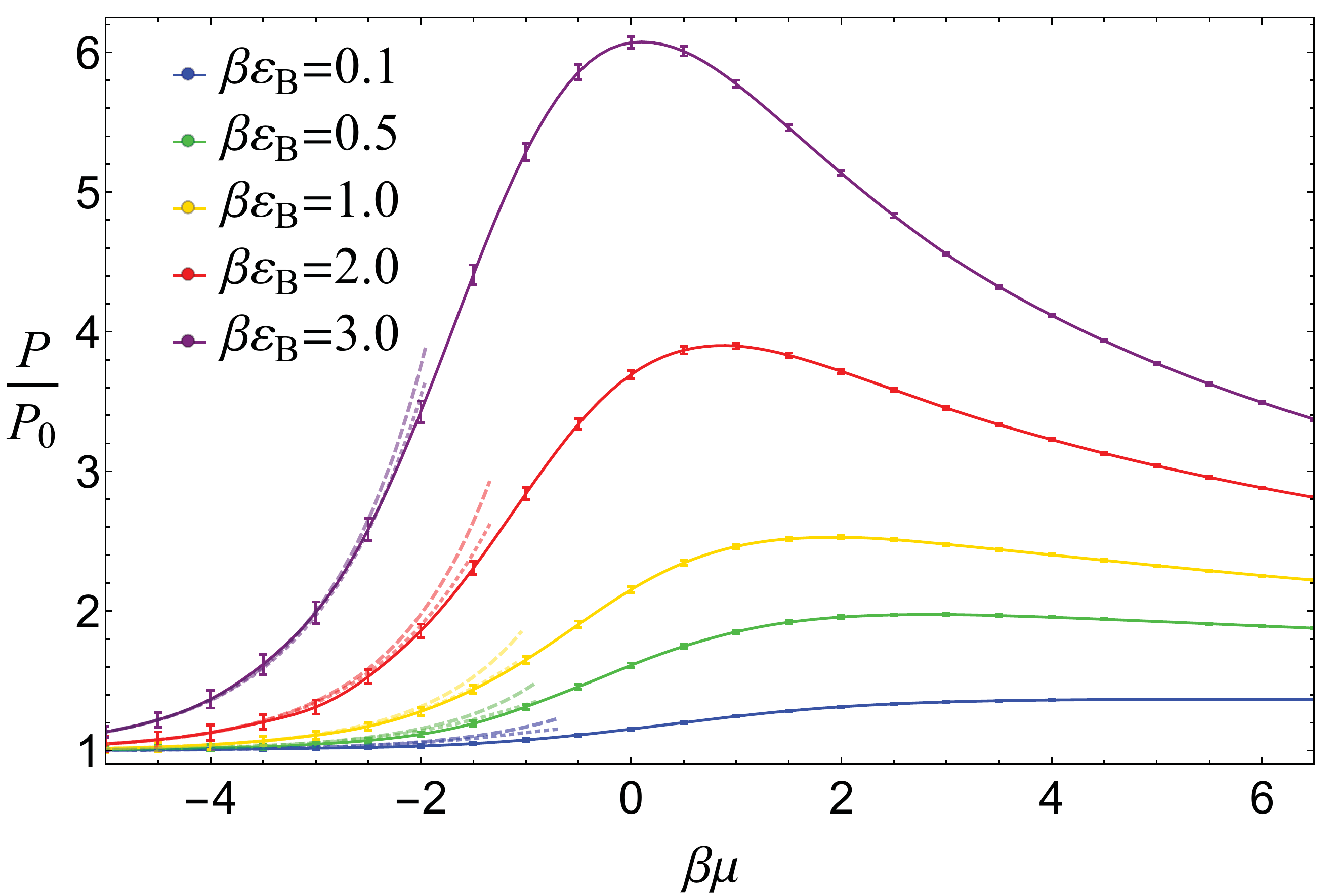}
\caption{\label{Fig:P_P0}(color online)
Pressure, in units of the non-interacting pressure $P^{}_0$, of spin-$1/2$ fermions in 2D for coupling strengths 
$\beta\varepsilon^{}_\text{B}$ = 0.1, 0.5, 1, 2, 3 (from bottom to top), as a function of $\beta \mu$. The error bars reflect the statistical uncertainty.
The long- and short-dashed lines show the second- and third-order virial expansion result, respectively.}
\end{figure}

Our calculations, as shown in Fig.~\ref{Fig:n_n0}, display a clear deviation from the Luttinger-Ward approach of 
Ref.~\cite{Enss2D} for $-2 \leq \beta \mu \leq 2$. This is not unexpected, as that regime is challenging for most
many-body approaches: it is the regime where quantum fluctuations are the strongest (at fixed $\beta \varepsilon^{}_\text{B}$).  
Away from that region, however, the agreement with Ref.~\cite{Enss2D} is satisfactory. Moreover, both calculations
seem to heal together to the virial expansion result at large and negative $\beta\mu$.
The starting point for this expansion is a Taylor expansion of the grand thermodynamic potential $\Omega$
in powers of the fugacity $z$,
\beq
\label{Eq:Omega}
- \beta \Omega =  Q_1^{} \left(z + b^{}_2 z^2 + b^{}_3 z^3 + \dots \right),
\eeq
where $Q_1^{} = 2 L^2 / \lambda^{2}_T$ is the 2D single-particle partition function and $b_n^{}$ are the virial
coefficients, which in general will be functions of the coupling $\beta \varepsilon^{}_\text{B}$.
Thus, the virial expansion for the density reads
%
$
{n \lambda^{2}_T}/{2} = z + 2 b_2^{} z^2 + 3 b_3^{} z^3 + \cdots 
$,
%
where the factor of $1/2$ on the left-hand side comes from the number of fermion species. 
For the system studied here, the second-order coefficient $b_2^{}$ is known analytically 
from the exact solution of the two-body problem (see e.g. Refs.~\cite{ChaffinSchaefer,ParishEtAl}):
\beq
\label{Eq:b2Exact}
b_2^{} = -\frac{1}{4} + e^{\beta \varepsilon^{}_\text{B}} -
\int_0^{\infty} \frac{d y}{y} \frac{2 e^{-\beta \varepsilon^{}_\text{B}  y^2}}{\pi^2 + 4 \ln^2 y}
.
\eeq
A calculation of $b_3^{}$ can be found in Ref.~\cite{ParishEtAl}.

To calculate the pressure $P$, we integrate as follows
\beq
\label{Eq:PIntegration}
P\lambda_T^4 = 2\pi \int_{-\infty}^{\beta \mu} {n \lambda^{2}_T} \; d (\beta \mu)',
\eeq
where we have put everything in dimensionless form using the thermal wavelength scale.
In Fig.~\ref{Fig:P_P0} we show $P$, as obtained from the above formula, in units of the pressure
of the non-interacting system $P^{}_0 \lambda_T^4 = 8 \pi I^{}_0(z)$. The virial expansion
of Eq.~(\ref{Eq:Omega}) is used in this integration to complete the approach to the $z\to0$ limit.

\begin{figure}[t]
\includegraphics[width=1.0\columnwidth]{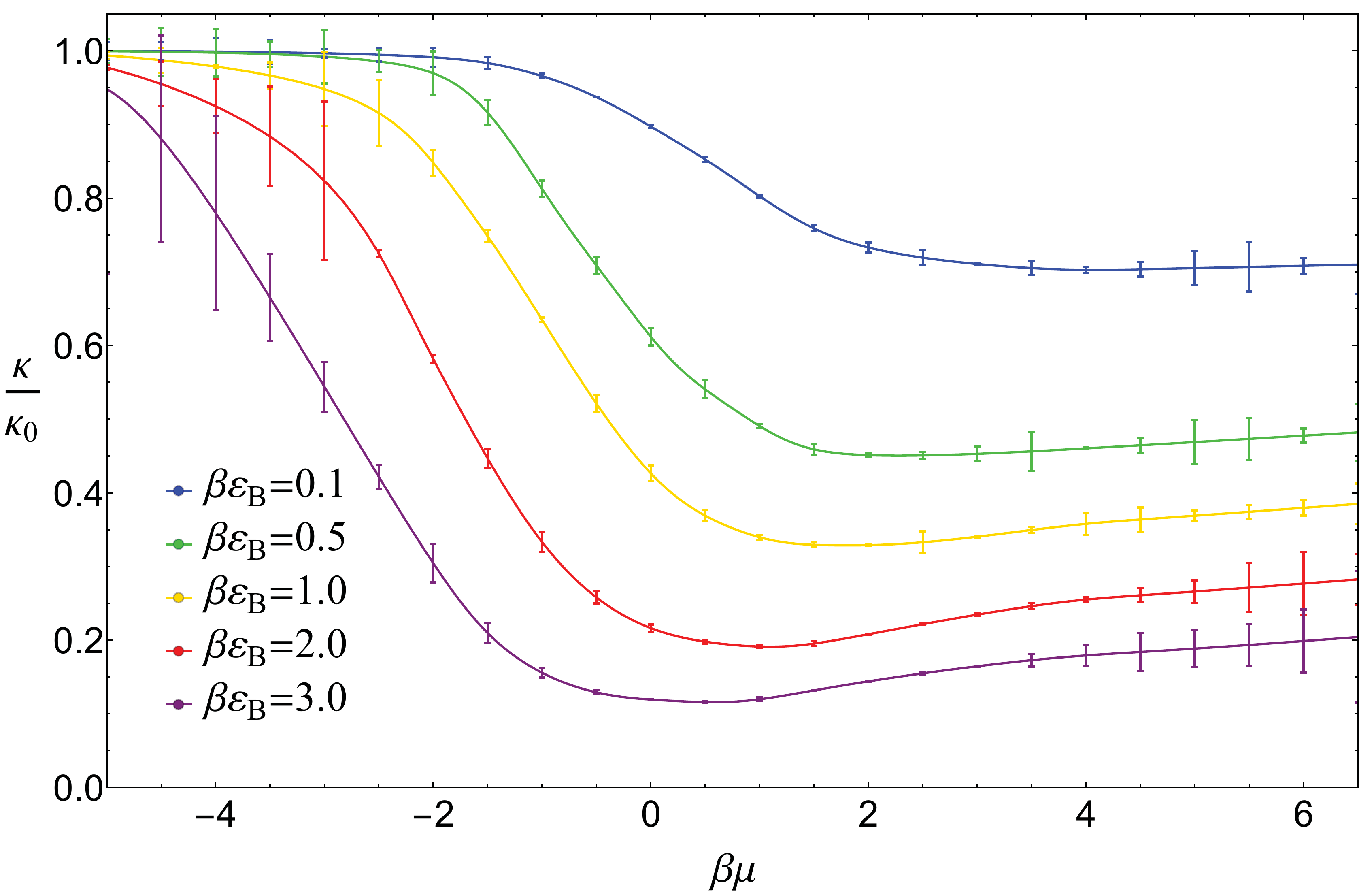}
\caption{\label{Fig:kappa}(color online)
Compressibility, in units of the non-interacting compressibility $\kappa^{}_0$, of spin-$1/2$ fermions in 2D for 
coupling strengths $\beta\varepsilon^{}_\text{B}$ = 0.1, 0.5, 1, 2, 3 (from top to bottom), as a function of $\beta \mu$.
The error bars reflect the difference between a smooth interpolation and the raw Monte Carlo data.
}
\end{figure}

On the other hand, by taking a derivative of $n$ one obtains the isothermal compressibility,
\beq
\kappa = \frac{\beta}{n^2}\left . \frac{\partial n}{\partial (\beta \mu)} \right |^{}_\beta = 
\frac{\lambda^{4}_T}{2\pi} \frac{1}{(n \lambda^{2}_T)^2} \left . \frac{\partial (n \lambda^{2}_T)}{\partial (\beta \mu)} \right |^{}_\beta .
\eeq
We report this quantity in Fig.~\ref{Fig:kappa}, in units of its non-interacting counterpart $\kappa^{}_0$, where
(in dimensionless form) $\kappa^{}_0 \lambda_T^{-4} = (2/\pi) (n^{}_0 \lambda^{2}_T)^{-2} I_2(z)$,
and $I^{}_2(z) = z\,{d I^{}_1(z)}/{d z}$.

To calculate the contact, we use the grand-canonical definition~\cite{ParishEtAl}
\beq
\label{Eq:Cdefinition}
C  \equiv \frac{2\pi}{\beta} \left .\frac{\partial (\beta \Omega)}{\partial \ln (a_\text{2D}^{}/\lambda_T^{})} \right|_{T,\mu}.
\eeq
From here, it is easy to see that the virial expansion for $C$ takes the form
\beq
\label{Eq:BetaContact}
\beta {C} = {2\pi}  Q_1^{} \left(c^{}_2 z^2 + c^{}_3 z^3 + \dots \right),
\eeq
where 
%
$
c^{}_n = -{\partial b_n^{}}/{\partial \ln (a_\text{2D}^{}/\lambda_T^{})}
$.
%
Using Eq.~\ref{Eq:b2Exact}, it is straightforward to obtain the exact continuum-limit answer:
\beq
\label{Eq:c2Exact}
c_2^{} = 2 \beta \varepsilon^{}_\text{B} e^{\beta \varepsilon^{}_\text{B}}\left[ 1 + 2 \int_0^{\infty} dy \frac{ y\; e^{-\beta \varepsilon^{}_\text{B} (y^2 + 1)}}{\pi^2 + 4 \ln^2 y} \right]
.
\eeq
This result was evidently used in Ref.~\cite{ParishEtAl}, but the formula itself is not shown explicitly anywhere, to the best of our knowledge.

\begin{figure}[t]
\includegraphics[width=1.0\columnwidth]{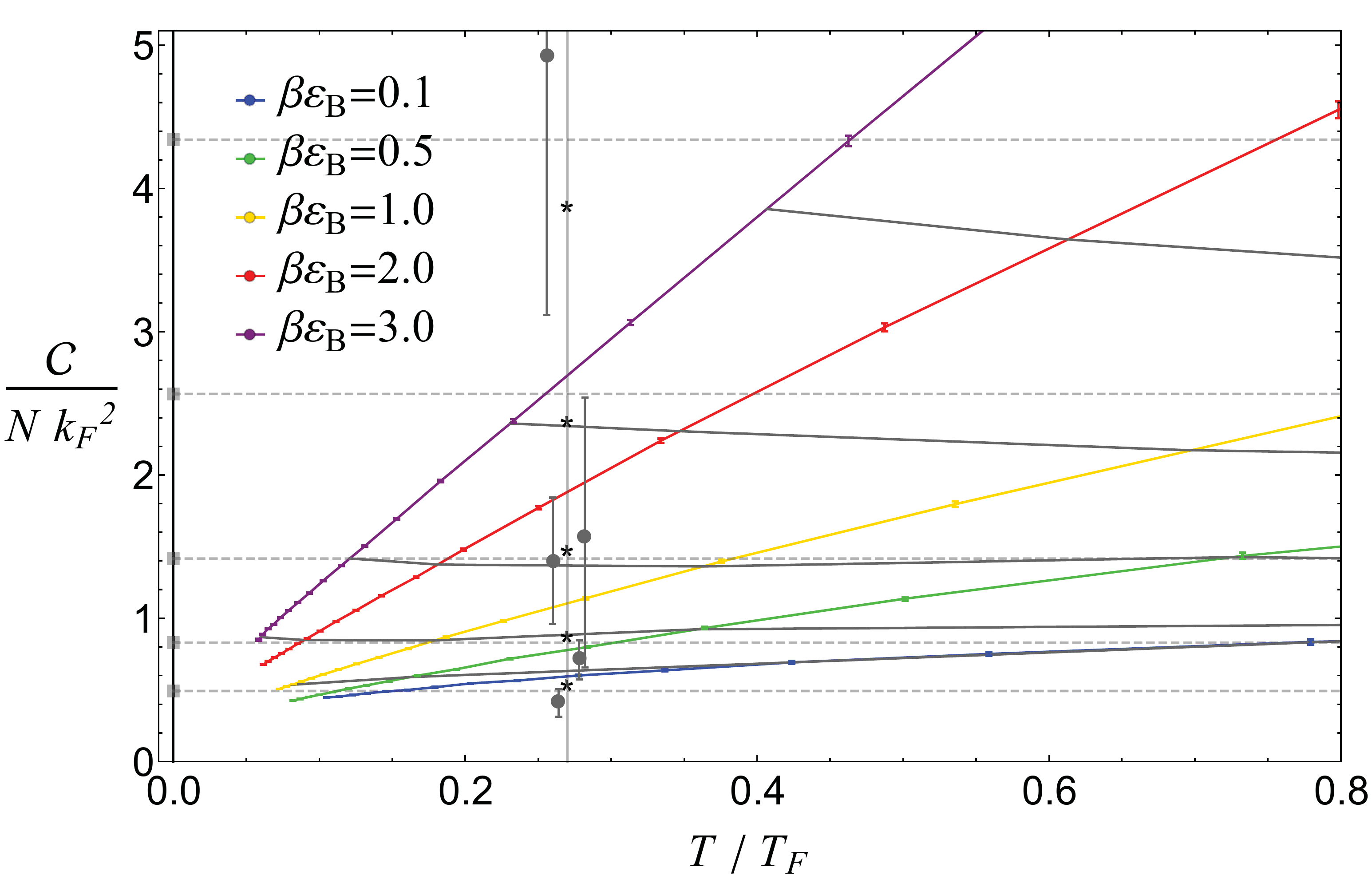}
\caption{\label{Fig:Contact}(color online)
Contact, in units of $1/(N k_{F}^2)$, for spin-$1/2$ fermions in 2D for coupling strengths $\beta\varepsilon^{}_\text{B}$ = 0.1, 0.5, 1, 2, 3 (from bottom to top), 
as a function of $T/T_F$. The error bars reflect the statistical uncertainty.
Solid lines of gray across colored contact results indicate lines of constant $\ln(k^{}_F {a}_\text{2D})=1.60,1.21,0.85,0.53,0.24$ from bottom to top.  
Corresponding experimental data at \(T/T_F=0.27\) from Ref.~\cite{ContactExperiment2D2012} given by solid circles (displaced about light gray line for visibility of error bars);
corresponding Luttinger-Ward results at \(T/T_F=0.27\) from Ref.~\cite{Enss2D} are shown as black stars;
and corresponding QMC calculations from Ref.~\cite{Bertaina} at \(T/T_F=0\) given by solid squares and dotted gray lines.}
\end{figure}

In our Monte Carlo calculations, we determine the contact by calculating the expectation value of the interaction energy $\hat V$.
Using the definition of Eq.~\ref{Eq:Cdefinition}, along with $- \beta \Omega = \ln \mathcal Z$, where $\mathcal Z$ is the
grand-canonical partition function, we obtain
\beq
C = \frac{-2\pi}{\beta} \frac{\partial \ln \mathcal Z}{\partial \ln (a_\text{2D}^{}/\lambda_T^{})} = 
- 2\pi \langle \hat V \rangle \frac{\partial \ln g}{\partial \ln (a_\text{2D}^{}/\lambda_T^{})},
\eeq
where we have used the fact that $\hat V$ is a contact interaction as in Eq.~\ref{Eq:H}. The remaining factor on the right is
determined by solving the 2-body problem. To turn this into an intensive, dimensionless quantity, 
we present results in the form $ C / (N k_{F}^2)$.
In Fig.~\ref{Fig:Contact} we show our lattice Monte Carlo results for this quantity as a function of $T/T_F$ and the 
coupling $\beta \varepsilon^{}_\text{B}$, as well as selected lines of constant $\ln(k^{}_F {a}_\text{2D})$. Corresponding to the latter, we have included ground-state quantum 
Monte Carlo (QMC) results \cite{Bertaina} and experimental results at finite temperature of  $T/T_F=0.27$ \cite{ContactExperiment2D2012}.  The experimental results largely agree 
with our calculations, however, the experimental errors and the maximum coupling calculated limit us from drawing any strong conclusions.  On the other hand, we note 
that the contact is largely flat at constant $\ln(k^{}_F {a}_\text{2D})$, which agrees qualitatively with
the Luttinger-Ward approach of Ref.~\cite{Enss2D}. We regard all this as evidence that the ground-state QMC results (see also next section) are in very good agreement 
with our finite-$T$ calculations, as the latter seem to approach the $T=0$ results in that limit.

Our full set of contact calculations as a function of $\beta \mu$ can be found in the Supplemental Material.  We have also verified there the agreement with the continuum-limit second-order virial expansion result (using Eq.~\ref{Eq:c2Exact}) in the high-energy regime. Removing lattice effects in this regime is
most demanding, which suggests that such systematic effects are well under control. 
%

{\it Additional Tests and Crosschecks.--~} 
%
\begin{table}[b]
\begin{center}
\caption{\label{Table:EGS}
Ground-state energy in units of its non-interacting counterpart, as a function of the coupling 
$\ln(k^{}_F {\tilde a}^{}_\text{2D})$. Here, ${\tilde a}^{}_\text{2D} = a^{}_\text{2D} e^\gamma/2$, where
$\gamma \!\simeq\! 0.577$ is Euler's constant, which is the convention followed in Ref.~\cite{Bertaina}.
}
\begin{tabularx}{\columnwidth}{@{\extracolsep{\fill}}c c c c}
\hline\hline
$\ln(k^{}_F {\tilde a}^{}_\text{2D})$ & $E^{}_\text{GS}/E^{}_\text{FG}$: Ref.~\cite{Bertaina} & This work & Difference\\
\hline
-2.00 & $-137.761(7)$ 	& $-139.4(5)$ 	& $1\%$	\\ 
-1.50 & $-50.593(4)$		& $-51.0(8)$ 		& $1\%$	\\
-1.00 & $-18.532(4)$		& $-19.0(5)$  	& $3\%$	\\
-0.50 & $-6.714(4)$		& $-6.8(2)$   		& $1\%$	\\
0.00  & $-2.318(2)$		& $-2.40(5)$   		& $3\%$	\\
0.25  & $-1.283(12)$		& $-1.35(3)$   	& $5\%$	\\
0.50  & $-0.638(10)$		& $-0.70(2)$   		& $10\%$	\\
1.44  & $0.349(6)$		& $0.38(1)$   	& $9\%$	\\
3.34  & $0.706(2)$  		& $0.698(1)$ 	& $1\%$	\\
\hline\hline
\end{tabularx}
\end{center}
\end{table}
%
As a test, we used the same lattice approach, in combination with the projection Monte Carlo method, to calculate
the ground-state energies of the system as a function of $\ln(k^{}_F a^{}_\text{2D})$, which were first determined 
non-perturbatively in Ref.~\cite{Bertaina}. Our results, along with those of Ref.~\cite{Bertaina},
are shown in Table~\ref{Table:EGS}.
Note that Ref.~\cite{Bertaina} uses $\ln(k^{}_F {\tilde a}^{}_\text{2D})$ as the coupling, where 
${\tilde a}^{}_\text{2D} = a^{}_\text{2D} e^\gamma/2$ and $\gamma \!\simeq\! 0.577$ is Euler's constant. 

{\it Summary and conclusions.--~} 
Using lattice Monte Carlo methods, we have calculated the finite temperature thermodynamics of homogeneous 
two-dimensional spin-$1/2$ fermions with attractive short-range interactions. We have presented results for the
density, pressure, compressibility, and Tan's contact for a wide range of temperatures 
and coupling strengths. Within our statistical and systematic uncertainties, our prediction for the density equation of state
differs from the prediction by Luttinger-Ward theory in a substantial region of parameter space. The
general agreement is nonetheless exceptional. We have also compared our calculations of the density and pressure 
with the second- and third-order virial expansion, with which they agree remarkably well in the low fugacity regime.
Moreover, the agreement seems stronger with our results than with the Luttinger-Ward approach.
Finally, we have presented a comparison of our calculation of the contact with previous ground-state 
calculations and finite-temperature experimental data. A more complete representation of our data for the contact, including
a comparison with the second-order virial expansion and an alternative temperature scale, appear in the Supplemental Material.

Our results for the density, pressure, and compressibility can also be compared with experiments. 
One of the motivations for the latter is that attractively interacting fermions in 2D are expected to undergo 
a Kosterlitz-Thouless transition into a superfluid phase at low enough temperatures. We do not see, in the 
quantities studied here, any particular signature of the transition. We defer further calculations in that direction 
to future work.


{\it Acknowledgements.--~}
We acknowledge useful discussions with
T. Enss, J. Levinsen, M. M. Parish, T. Sch\"afer, J. E. Thomas, and C. J. Vale, as well as
M. D. Hoffman, P. D. Javernick, A. C. Loheac, and W. J. Porter.
We thank M. Bauer, M. M. Parish, and T. Enss for
sharing their results for our Fig.~\ref{Fig:n_n0}, 
and M. M. Parish and J. Levinsen for their results on the third-order virial coefficient.
This material is based upon work supported by the National Science Foundation 
Nuclear Theory Program under Grant No. PHY{1306520}.



\section{Supplemental online material }

We display our data for the temperature and contact in scales different from those in the main text.
We compare the contact explicitly with the second-order virial expansion.

\section{Temperature scale}

\begin{figure}[h]
\includegraphics[width=1.0\columnwidth]{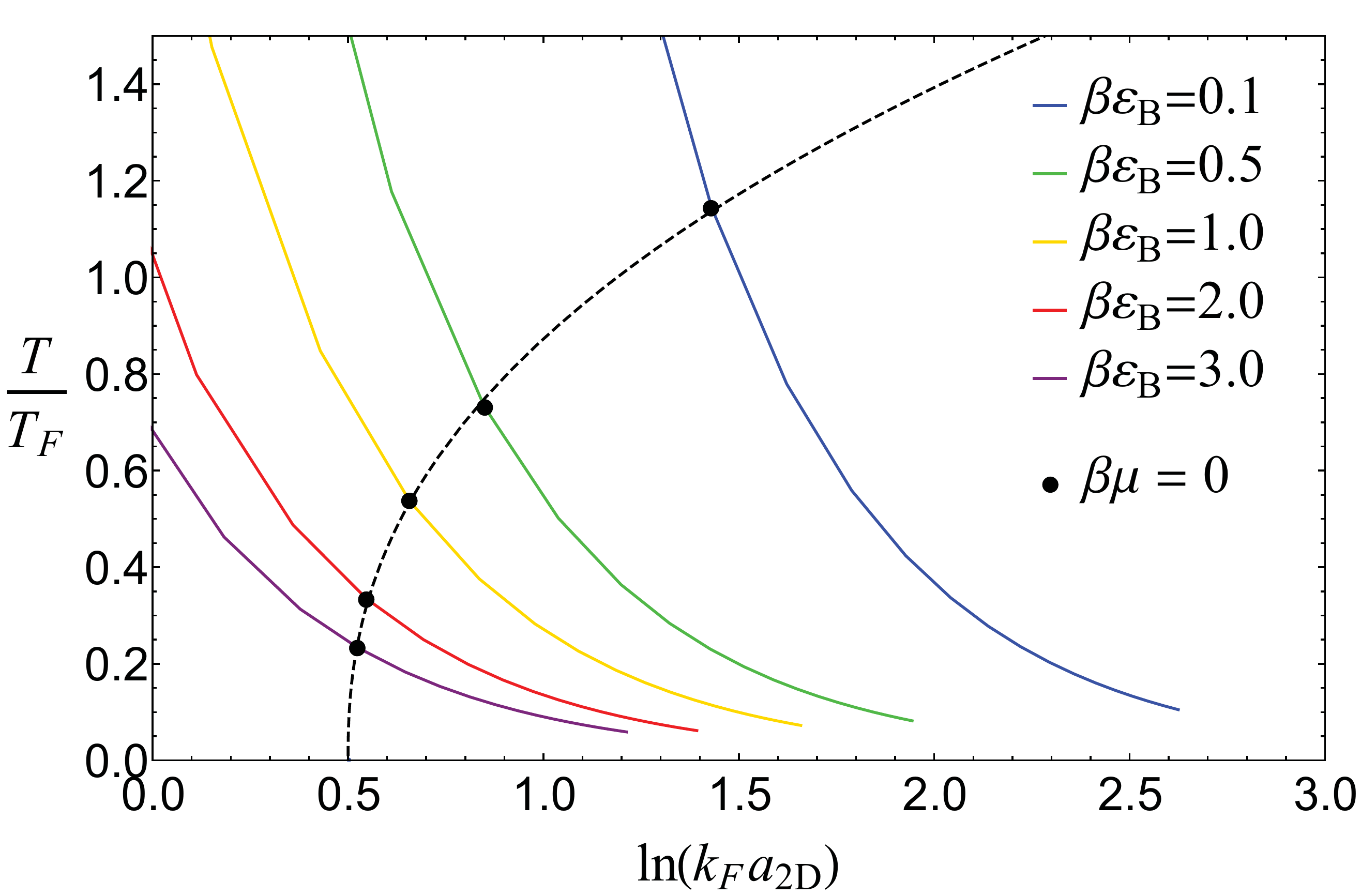}
\caption{\label{Fig:T_eF}(color online)
Temperature scale $T$, in units of the non-interacting Fermi energy $\varepsilon^{}_F$, of spin-$1/2$ fermions in 2D, 
for coupling strengths $\beta\varepsilon^{}_\text{B}$ = 0.1, 0.5, 1, 2, 3 (from top to bottom), as a function of the more common coupling
strength parameter $\ln(k^{}_F a^{}_\text{2D})$. The dashed line is a fit to the $\beta\mu=0$ points for each coupling; 
the fit form is $A( \ln(k^{}_F a^{}_\text{2D}) - 0.5)^{B}$, with $A = 1.170$ and $B = 0.422$. 
The value at $T=0$, namely $\ln(k^{}_F a^{}_\text{2D})\simeq0 .5$, was estimated from Ref.~\cite{Bertaina2}.
}
\end{figure}

Although $\beta \mu$ is a useful dimensionless parameter to characterize the temperature, especially in the grand-canonical
ensemble, it is often even more useful to present the temperature in a scale set by the density, namely the Fermi energy 
$\varepsilon^{}_F = k^{2}_F/2$, where $k_F^{} = \sqrt{2\pi n}$ is the Fermi momentum.
In Fig.~\ref{Fig:T_eF} we show $T/\varepsilon^{}_F$ as a function of $\ln (k^{}_F a_\text{2D}^{})= \ln (2\varepsilon^{}_F/\varepsilon^{}_\text{B})/2$, 
where $a_\text{2D}^{} = 1/\sqrt{\varepsilon^{}_\text{B}}$. The statistical uncertainties are not shown, but from the smoothness of
the resulting curves it can be inferred that they are very small. Their size results from the tiny uncertainty in the density $n$ 
(see density plot in main text), which (barely) impacts both axes in Fig.~\ref{Fig:T_eF} .

%

\section{Tan's contact}

\begin{figure}[t]
\includegraphics[width=1.0\columnwidth]{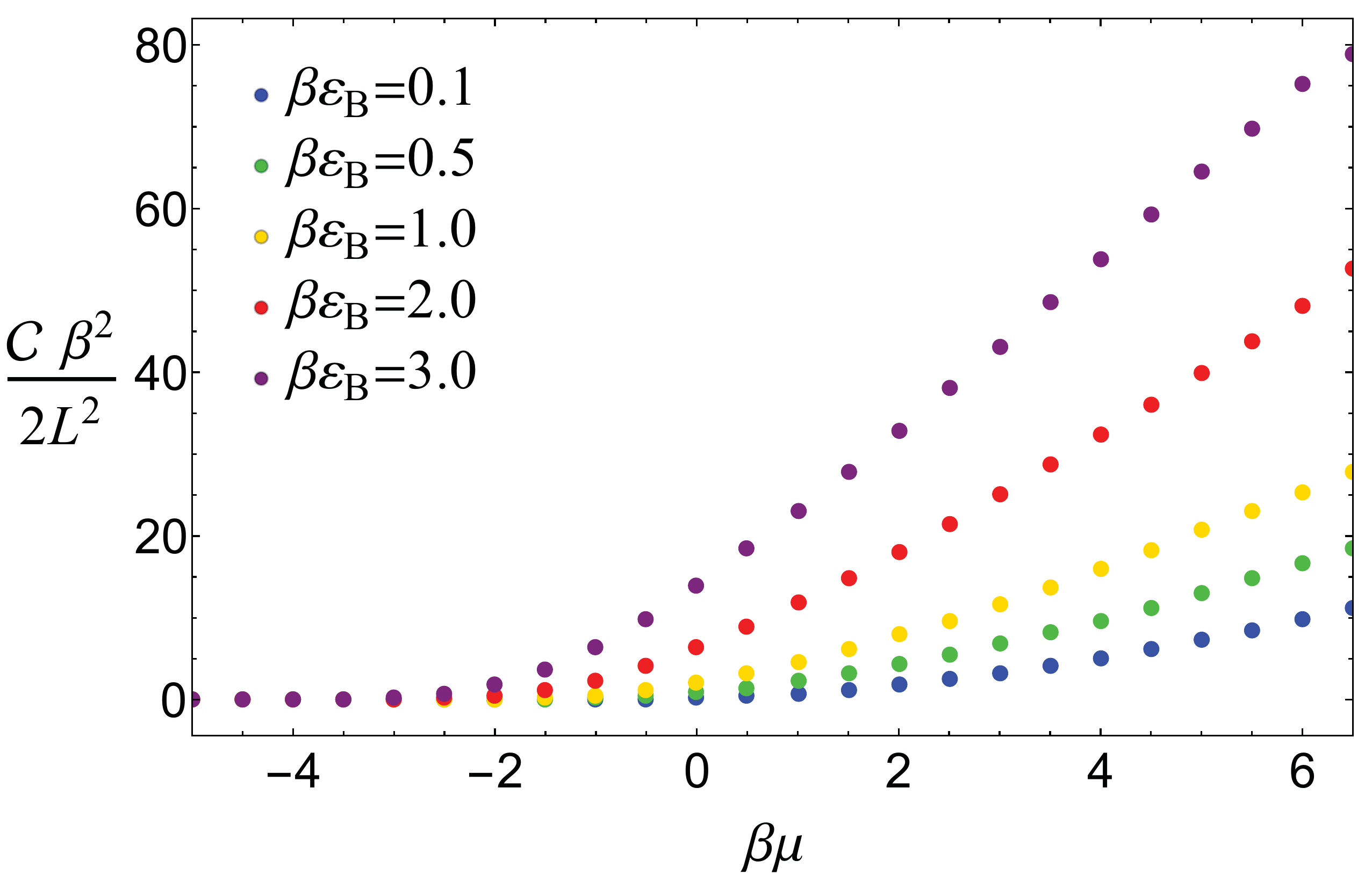}
\caption{\label{Fig:C1}(color online) Contact of spin-$1/2$ fermions in 2D, for coupling strengths 
$\beta\varepsilon^{}_\text{B}$ = 0.1, 0.5, 1, 2, 3 (from bottom to top), as a function of $\beta \mu$.}
\end{figure}

In the main text we showed Tan's contact $C$ as a function of $T/\varepsilon^{}_F$. However, in our calculations $\varepsilon^{}_F$ is
computed as an output after the average density is determined. Therefore, it is more natural to report $C$ as a function
of $\beta \mu$, which we do in Fig.~\ref{Fig:C1}. The statistical uncertainties in our Monte Carlo data are smaller than the size of the symbols
in that figure.

\begin{figure}[b]
\includegraphics[width=1.0\columnwidth]{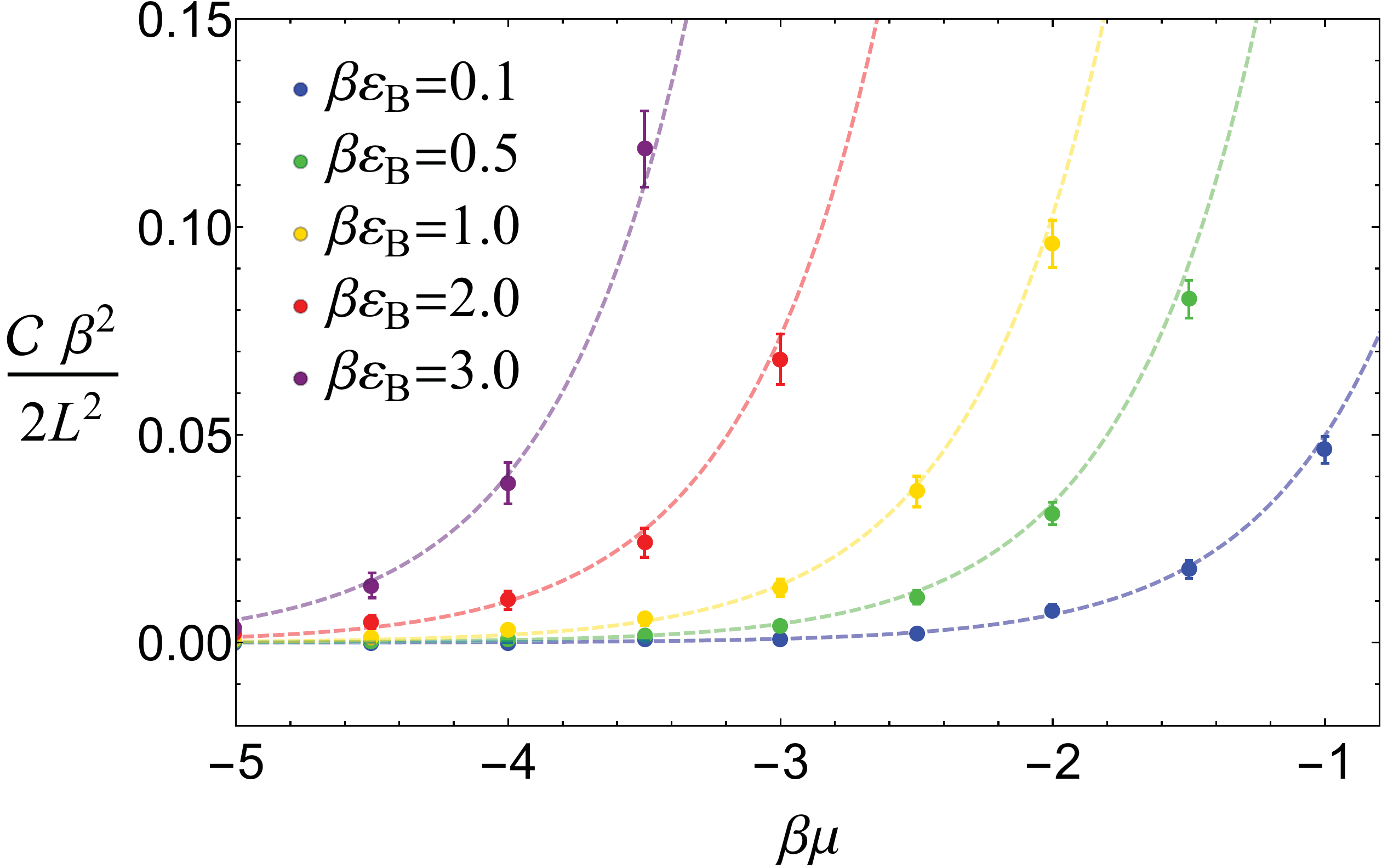}
\caption{\label{Fig:C2}(color online) Contact of spin-$1/2$ fermions in 2D, for coupling strengths 
$\beta\varepsilon^{}_\text{B}$ = 0.1, 0.5, 1, 2, 3 (from bottom to top), as a function of 
$\beta \mu$ in the low-fugacity regime. The dashed lines show the second-order virial expansion result.}
\end{figure}

%
An additional benefit of the above representation is that it allows for a more direct comparison with the virial expansion at
large and negative $\beta \mu$. Our calculations agree with that expansion to second order (leading order for the contact),
as can be appreciated in Fig.~\ref{Fig:C2}. The statistical uncertainties can be seen in this case.

\begin{figure}[t]
\includegraphics[width=1.0\columnwidth]{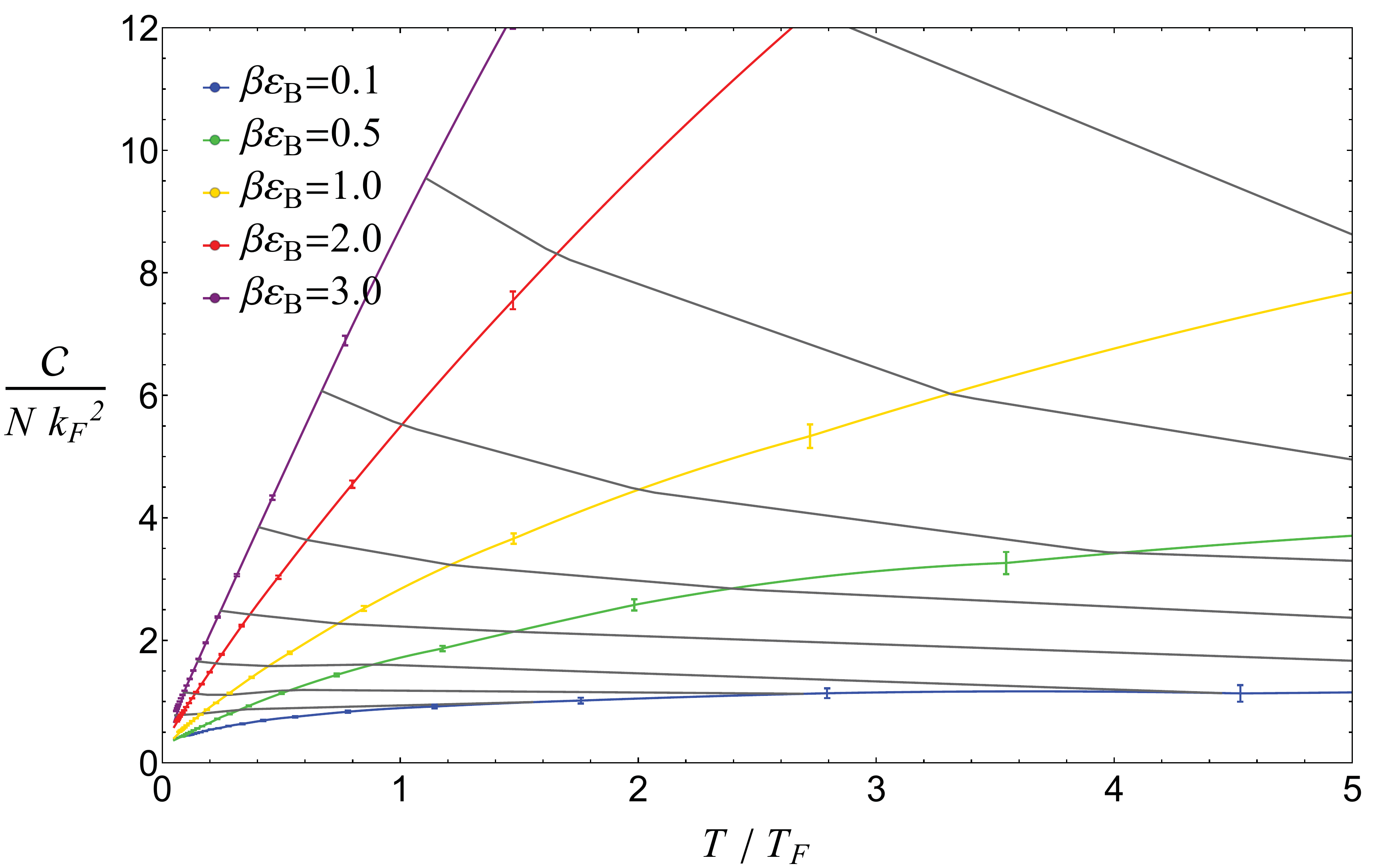}
\caption{\label{Fig:C3}(color online) Contact of spin-$1/2$ fermions in 2D as a function of 
$\beta \mu$. The solid colored lines show lines of constant $\beta\varepsilon^{}_\text{B}$ = 0.1, 0.5, 1, 2, 3 (from bottom to top). 
Solid lines of gray across colored contact results indicate lines of constant $\ln(k^{}_F {a}_\text{2D})$=1.25, 1.00, 0.75, 0.5, 0.25, 0, -0.25, -0.5 
from bottom to top.
}
\end{figure}

%
%
Finally, to further complement the contact plots of the main text, we provide in Fig.~\ref{Fig:C3} the full temperature 
range we explored in our calculations.

\begin{thebibliography}{99}

\bibitem{BKT}
{\it 40 Years of Berezinskii-Kosterlitz-Thouless Theory}
J.V. Jose (Ed.) (World Scientific, Singapore, 2013);
J.M. Kosterlitz and D.J. Thouless, J. Phys. C {\bf 6}, 1181 (1973);
J.M. Kosterlitz, ibid. {\bf 7}, 1046 (1974).

\bibitem{confinement}
I. I. Kogan and A. Kovner,
{\it At the Frontier of Particle Physics}, Chapter 10, pp.2336.
(World Scientific, Singapore, 2003).

\bibitem{Hoffmann}
J. Hofmann,
Phys. Rev. Lett. {\bf 108}, 185303 (2012);
E. Taylor and M. Randeria,
Phys. Rev. Lett. {\bf 109}, 135301 (2012);
Phys. Rev. Lett. {\bf 110}, 089904 (2013);

\bibitem{HighTc}
D. N. Basov and T. Timusk,
Rev. Mod. Phys. {\bf 77}, 721 (2005);
S. A. Kivelson, I. P. Bindloss, E. Fradkin, V. Oganesyan, J. M. Tranquada, A. Kapitulnik, and C. Howald,
Rev. Mod. Phys. {\bf 75}, 1201 (2003)

\bibitem{graphene}
V. N. Kotov, B. Uchoa, V. M. Pereira, F. Guinea, and A. H. Castro Neto,
Rev. Mod. Phys. {\bf 84}, 1067 (2012).

\bibitem{RevExp}
\textit{Ultracold Fermi Gases}, 
Proceedings of the International School of Physics ``Enrico Fermi", Course CLXIV, 
Varenna, June 20 -- 30, 2006, 
M.~Inguscio, W.~Ketterle, C.~Salomon (Eds.) (IOS Press, Amsterdam, 2008).


\bibitem{RevTheory}
I.~Bloch, J.~Dalibard, and W.~Zwerger,
Rev. Mod. Phys. {\bf 80}, 885 (2008);
S. Giorgini, L. P. Pitaevskii, and S. Stringari,
Rev. Mod. Phys. \textbf{80}, 1215 (2008).


\bibitem{Experiments2D2010}
K. Martiyanov, V. Makhalov, and A. Turlapov, Phys. Rev. Lett. {\bf 105}, 030404 (2010).


\bibitem{Experiments2D2011}
M. Feld, B. Fr\"ohlich, E. Vogt, M. Koschorreck, and M. K\"ohl, Nature (London) {\bf 480}, 75 (2011);
B. Fr\"ohlich, M. Feld, E. Vogt, M. Koschorreck, W. Zwerger, and M. K\"ohl, Phys. Rev. Lett. {\bf 106}, 105301 (2011);
P. Dyke, E.D. Kuhnle, S. Whitlock, H. Hu, M. Mark, S. Hoinka, M. Lingham, P. Hannaford, and C. J. Vale, Phys. Rev. Lett. {\bf 106}, 105304 (2011);
A. A. Orel, P. Dyke, M. Delahaye, C. J. Vale, and H. Hu, New J. Phys. {\bf 13}, 113032 (2011).


\bibitem{Experiments2D2012}
A. T. Sommer, L. W. Cheuk, M. J. H. Ku, W. S. Bakr, and M. W. Zwierlein, Phys. Rev. Lett. {\bf 108}, 045302 (2012);
Y. Zhang, W. Ong, I. Arakelyan, and J. E. Thomas, Phys. Rev. Lett. {\bf 108}, 235302 (2012);
S.K. Baur, B. Fr\"ohlich, M. Feld, E. Vogt, D. Pertot, M. Koschorreck, and M. K\"ohl, Phys. Rev. A {\bf 85}, 061604 (2012);
M. Koschorreck, D. Pertot, E. Vogt, B. Frohlich, M. Feld, and M. Kohl, Nature (London) {\bf 485}, 619 (2012);
E. Vogt, M. Feld, B. Fr\"ohlich, D. Pertot, M. Koschorreck, M. K\"ohl, Phys. Rev. Lett. {\bf 108}, 070404 (2012).

\bibitem{ContactExperiment2D2012}
B. Fr\"ohlich, M. Feld, E. Vogt, M. Koschorreck, M. K\"ohl, C. Berthod, and T. Giamarchi, Phys. Rev. Lett. {\bf 109}, 130403 (2012).


\bibitem{RanderiaPairingFlatLand}
M. Randeria, Physics {\bf 5}, 10 (2012).


\bibitem{Experiments2D2014}
V. Makhalov, K. Martiyanov, A. Turlapov, Phys. Rev. Lett. {\bf 112}, 045301 (2014).


\bibitem{Vale2Dcriteria}
P. Dyke, K. Fenech, T. Peppler, M. G. Lingham, S. Hoinka, W. Zhang, B. Mulkerin, H. Hu, X.-J. Liu, C. J. Vale,
arXiv:1411.4703.


\bibitem{Miyake}
K. Miyake, 
Prog. Theor. Phys. {\bf 69}, 1794 (1983).


\bibitem{BCSBEC2D}
M. Randeria, J.-M. Duan, and L.-Y. Shieh, 
Phys. Rev. Lett. {\bf 62}, 981 (1989); 
Phys. Rev. B {\bf 41}, 327 (1990);
S. Schmitt-Rink, C.M. Varma, and A.E. Ruckenstein,
Phys. Rev. Lett. {\bf 63}, 445 (1989); 
M. Drechsler and W. Zwerger, 
Ann. Phys. (Leipzig) {\bf 1}, 15 (1992).


\bibitem{ZhangLinDuan}
W. Zhang, G.-D. Lin, and L.-M. Duan,
Phys. Rev. A {\bf 77}, 063613 (2008).


\bibitem{Bertaina}
G. Bertaina and S. Giorgini,
Phys. Rev. Lett. {\bf 106}, 110403 (2011).


\bibitem{ShiChiesaZhang}
H. Shi, S. Chiesa, and S. Zhang,
Arxiv:1504.00925.


\bibitem{LiuHuDrummond}
X.-J. Liu, H. Hu, and P. D. Drummond, 
Phys. Rev. B {\bf 82}, 054524 (2010).


\bibitem{Enss2D}
M. Bauer, M. M. Parish, and T. Enss,
Phys. Rev. Lett. {\bf 112}, 135302 (2014).


\bibitem{ParishEtAl}
V. Ngampruetikorn, J. Levinsen, and M. M. Parish,
Phys. Rev. Lett. {\bf 111}, 265301 (2013).


\bibitem{BarthHofmann}
M. Barth and J. Hofmann,
Phys. Rev. A {\bf 89}, 013614 (2014).


\bibitem{ChaffinSchaefer}
C. Chaffin and T. Sch\"afer
Phys. Rev. A {\bf 88}, 043636 (2013).

\bibitem{EnssUrban}
S. Chiacchiera, D. Davesne, T. Enss, and M. Urban,
Phys. Rev. A {\bf 88}, 053616 (2013).

\bibitem{BaurVogt}
S. K. Baur, E. Vogt, M. K\"ohl, and G. M. Bruun,
Phys. Rev. A 87, 043612 (2013).

\bibitem{EnssShear}
T. Enss, C. K\"uppersbusch, L. Fritz,
Phys. Rev. A {\bf 86}, 013617 (2012).


\bibitem{TanContact}
S. Tan, Ann. Phys. {\bf 323}, 2952 (2008);
{\it ibid.} {\bf 323}, 2971 (2008);
{\it ibid.} {\bf 323}, 2987 (2008);
S. Zhang, A. J. Leggett, 
Phys. Rev. A {\bf 77}, 033614 (2008);
F. Werner, 
Phys. Rev. A {\bf 78}, 025601 (2008);
E. Braaten, L. Platter, 
Phys. Rev. Lett. {\bf 100}, 205301 (2008);
E. Braaten, D. Kang, L. Platter,
{\it ibid.} {\bf 104}, 223004 (2010);
C. Langmack, M. Barth, W. Zwerger, E. Braaten,
Phys. Rev. Lett. {\bf 108}, 060402 (2012).
J.E. Drut, T.A. L\"{a}hde, T. Ten, 
Phys. Rev. Lett. {\bf 106}, 205302 (2011);
K. Van Houcke, F. Werner, E. Kozik, N. Prokof'ev, B. Svistunov,
arXiv:1303.6245.


\bibitem{ContactReview}
F. Werner and Y. Castin, Phys. Rev. A {\bf 86}, 013626 (2012).
E. Braaten, 
in {\it The BCS-BEC Crossover and the Unitary Fermi Gas}, 
edited by W. Zwerger (Springer-Verlag, 2012).
X.-J. Liu
Phys. Rep. {\bf 524}, 37 (2013).


\bibitem{WernerCastin}
F. Werner and Y. Castin, Phys. Rev. A {\bf 86}, 013626 (2012).


\bibitem{EoS1D}
M. D. Hoffman, P. D. Javernick, A. C. Loheac, W. J. Porter, E. R. Anderson, J. E. Drut,
Phys. Rev. A {\bf 91}, 033618 (2015).


\bibitem{BDM1}
A.~Bulgac, J.~E. Drut, and P.~Magierski,
Phys. Rev. Lett. {\bf 96}, 090404 (2006).


\bibitem{BDM2}
A. Bulgac, J.~E. Drut, and P. Magierski, 
Phys. Rev. A {\bf 78}, 023625 (2008).


\bibitem{EoSUFG2}
J. E. Drut, T. A. L\"ahde, G. Wlazlowski, P. Magierski,
Phys. Rev. A {\bf 85}, 051601(R) (2012).


\bibitem{HS}
R.~L.~Stratonovich,
Sov.\ Phys.\ Dokl. {\bf 2}, 416 (1958);
J.~Hubbard,
Phys.\ Rev.\ Lett. {\bf 3}, 77 (1959).


\bibitem{Drut:2012md} 
J.~E.~Drut and A.~N.~Nicholson,
J.\ Phys.\ G {\bf 40}, 043101 (2013).


\bibitem{QMCReviews}
D.~Lee, Phys.\ Rev.\ C {\bf 78}, 024001 (2008);
Prog.\ Part.\ Nucl.\ Phys. {\bf 63}, 117 (2009).


\bibitem{ScatteringIn2D}
S. K. Adhikari, 
Am. J. Phys. {\bf 54}, 362 (1986).


\end{thebibliography}

\begin{thebibliography}{99}

\bibitem{Bertaina2}
G. Bertaina and S. Giorgini,
Phys. Rev. Lett. {\bf 106}, 110403 (2011).

\end{thebibliography}
\end{document}